\documentclass[singlecolumn]{article}
\usepackage{PRIMEarxiv}
\usepackage{cite}
\usepackage{amsmath,amssymb,amsfonts}
\usepackage{hyperref}       
\usepackage{amsfonts}       
\usepackage{fancyhdr}       
\usepackage{graphicx}       
\usepackage[squaren]{SIunits}
\usepackage{tikz}
\usepackage[percent]{overpic}
\usepackage{xcolor}

\title{3DMPR — A robust morphological approach for applying phase retrieval in proximity to highly-attenuating objects in CT}
\author{
 J. A. Pollock\textsuperscript{1}, L. C. P. Croton\textsuperscript{1}, K. S. Morgan\textsuperscript{1}, K. J. Crossley\textsuperscript{2,}\textsuperscript{3}, M. J. Wallace\textsuperscript{2,}\textsuperscript{3}, \\ \textbf{G. A. Buckley\textsuperscript{4}, S. B. Hooper\textsuperscript{2,}\textsuperscript{3}, M. J. Kitchen\textsuperscript{1}} \\
 \\
\textsuperscript{1}
School of Physics and Astronomy, Monash University, Clayton, VIC, Australia \\
\textsuperscript{2}
The Ritchie Centre, Hudson Institute of Medical Research, Clayton, VIC, Australia. \\
\textsuperscript{3}
Department of Obstetrics and Gynaecology, Monash University, Wellington Rd, Clayton, VIC, Australia. \\
\textsuperscript{4}
Walter and Eliza Hall Institute of Medical Research, Parkville, VIC, Australia
}

\begin{document}
\maketitle

\begin{abstract}
X-ray imaging is a fast, precise and non-invasive method of imaging which, when combined with computed tomography, provides detailed 3D rendering of samples. Incorporating propagation-based phase contrast can vastly improve data quality for weakly attenuating samples via phase retrieval, allowing radiation exposure to be reduced. However, applying phase retrieval to multi-material samples commonly requires choice of which material boundary to tune the reconstruction. Selecting the boundary with strongest phase contrast increases noise suppression, but at the detriment of over-blurring other interfaces and potentially removing quantitative sample information. Additionally, conventional phase-retrieval algorithms cannot be used for regions bounded by more than one material, requiring alternative methods. Here we present a computationally-efficient, non-iterative nor AI-mediated method for applying strong phase retrieval, whilst preserving sharp boundaries for all materials within the sample. 3D phase retrieval is combined with morphological operations to prevent over-blurring artefacts from being introduced, while avoiding the potentially long convergence times required by iterative approaches. This technique, entitled 3DMPR, was tested on phase contrast images of a rabbit kitten brain encased by the surrounding dense skull. Using \unit{24}{\kilo e \volt} synchrotron radiation with a \unit{5}{\meter} propagation distance, 3DMPR provided a 6.8-fold improvement in the signal-to-noise ratio (SNR) of brain tissue over the standard phase retrieval procedure, without over-smoothing the images. Simultaneous quantification of edge resolution and SNR gain was performed with an aluminium-water phantom imaged using a microfocus X-ray tube at \unit{35}{\kilo V_p} and \unit{0.576}{\meter} effective propagation distance. There, 3DMPR provided a 4-fold SNR boost whilst preserving the boundary spatial resolution at \unit{54\pm1}{\micro\meter}, compared to \unit{108\pm2}{\micro\meter} using conventional phase retrieval. These results illustrate the ability of 3DMPR to create new avenues of dose reduction in clinical settings. 
\end{abstract}

\section{Introduction}
\label{sec:Introduction}
X-ray imaging can provide high-resolution, three-dimensional (3D) visualisation of the internal structure of an object when combined with computed tomography (CT). However, capturing high-resolution CT images with high signal-to-noise ratio (SNR) using standard x-ray systems requires a non-negligible radiation dose, potentially harming samples that are sensitive to ionising radiation. Hospital imaging systems limit radiation exposure by using highly-efficient X-ray detectors, with relatively low resolution being the typical trade-off. Iterative reconstruction algorithms may also be used to help balance noise and spatial resolution \cite{vliegenthart_innovations_2022}. 

\begin{figure}
	\centering
	\begin{overpic}[width=0.75\columnwidth,,]{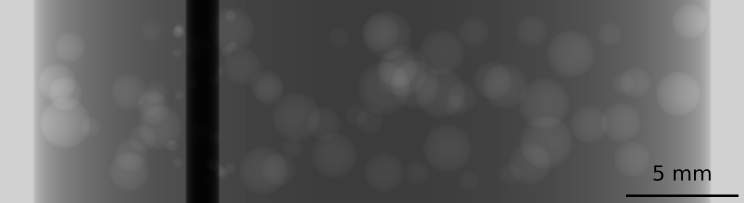}
    \put(0.5,22.0){\shortstack[l]{\fontsize{16}{18}\selectfont \textcolor{black}{(a)}}}
    \put(-10.0, 0.0){\rotatebox{90}{\shortstack[l]{\fontsize{14}{16}\selectfont \parbox{3cm}{\centering Exit \\ \vspace{-0.5mm} Surface}}}}
    \put(30.4, 5.4){
    \begin{tikzpicture}
      \draw[white, line width = 2.0pt, ->, >=to] (0.6,0.75) -- (0,0);
    \end{tikzpicture}}
    \end{overpic}\hfill%
	\begin{overpic}[width=0.75\columnwidth,,]{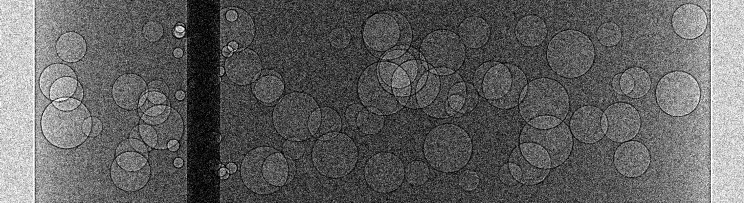}
    \put(0.5,22.0){\shortstack[l]{\fontsize{16}{18}\selectfont \textcolor{black}{(b)}}}
    \put(-10.0, 0.0){\rotatebox{90}{\shortstack[l]{\fontsize{14}{16}\selectfont \parbox{3cm}{\centering Phase \\ \vspace{-0.5mm} Contrast}}}}
    \end{overpic}\hfill%
	\begin{overpic}[width=0.75\columnwidth,,]{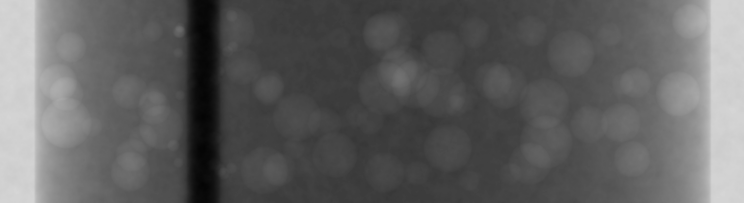}
    \put(0.5,22.0){\shortstack[l]{\fontsize{16}{18}\selectfont \textcolor{black}{(c)}}}
    \put(-10.0, 0.0){\rotatebox{90}{\shortstack[l]{\fontsize{14}{16}\selectfont \parbox{3cm}{\centering \vspace{-0.5mm} Single Material}}}}
    \end{overpic}\hfill%
	\begin{overpic}[width=0.75\columnwidth,,]{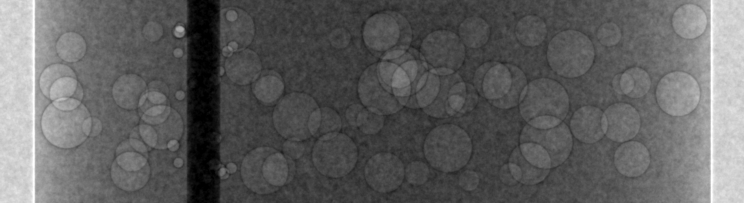}
    \put(0.5,22.0){\shortstack[l]{\fontsize{16}{18}\selectfont \textcolor{black}{(d)}}}
    \put(-10.0, 0.0){\rotatebox{90}{\shortstack[l]{\fontsize{14}{16}\selectfont \parbox{3cm}{\centering \vspace{-0.5mm} Two \\ Material}}}}
    \end{overpic}\\
    
	\caption{Simulations of phase-retrieval algorithms applied to projections of a two-material phantom. This phantom is composed of low-Z PMMA embedded with spherical cavities (white arrow) and a high-z aluminium cylinder which appears as a dark column in the exit surface wavefield shown in (a). (b) shows the phase contrast image produced from free-space propagation via the TIE to produce Fresnel fringes at each boundary that vary in amplitude according to the materials present.  Here, noise is added to simulate low-flux imaging. (c) Single material phase retrieval with Eq. \ref{equation: paganin}, using $\delta$ and $\mu$ for PMMA, shows accurate restoration of the low-Z/air boundaries (cf. (a)) but severely over-blurs the high-Z material and the surrounding area. This can be avoided with (d) two material phase retrieval to the high-Z, low-Z interface, using Eq. \ref{equation: beltran} with the $\delta$ and $\mu$ values for the low-Z (PMMA) and high-Z (Al) materials, but at the expense of higher remnant noise and phase contrast fringes left around the low-Z/air boundaries of each cavity. An ideal approach would achieve the benefits of both approaches shown in (c) and (d). }
	\label{figure:projectionsims}
\end{figure}

High-resolution CT scans of biological tissues can be achieved with low radiation dose by using phase contrast x-ray imaging \cite{kitchen_ct_2017, arhatari_x-ray_2021}. Phase contrast imaging uses specific setups to transform phase effects, introduced by the sample, into intensity variations capable of being recorded by detectors, and hence provides more information than from absorption contrast alone. While several x-ray phase contrast techniques exist, we focus here on propagation-based imaging (PBI), taking advantage of the simplicity in optical design that only requires a sufficiently-coherent\cite{pelliccia_theory_2017} x-ray wavefield at the sample position and some propagation distance between the sample and detector.  PBI has relaxed temporal coherence requirements and hence is applicable to macro- \cite{kitchen_ct_2017, pollock_low-dose_2024} and micro-scale \cite{bidola_application_2017}  imaging with synchrotron and laboratory sources. Phase contrast arises from interference between sections of the x-ray wavefield that have incurred different phase shifts as they pass through an object. This results in intensity fringes appearing along material boundaries at the detector plane, effectively acting as a sharpening filter \cite{gureyev_unreasonable_2017}. This is particularly valuable for those boundaries between low-Z materials, which are weakly-attenuating and hence create only weak attenuation contrast. To restore the sample structure requires the application of a phase-retrieval algorithm. For objects comprised of a single monomorphous material, the algorithm of Paganin et. al \cite{paganin_simultaneous_2002} accomplishes this by inverting the transport-of-intensity equation (TIE). The result is a low-pass filter specifically tuned to suppress phase contrast fringes at the boundaries of the object, whilst also providing SNR amplification by  suppressing high-frequency noise. The algorithm can be represented by
\begin{equation}
I(x, y) = \mathcal{F}^{-1}\left\{\frac{\mathcal{F}[I_d(x, y)/I_0]}{1 + \frac{\delta \Delta}{\mu}\textbf{k}_{\perp}^2}\right\} \label{equation: paganin},
\end{equation}
where $\mathcal{F}$ and $\mathcal{F}^{-1}$ are the forward and inverse Fourier transform operators, $I_d(x, y)$ is the image measured at the detector plane, $I_0$ is the spatially variable incident intensity, $\Delta$ is the propagation distance from object to detector, $\delta$ represents the real difference of the complex refractive index from unity and $\mu$ is the linear attenuation coefficient. $\textbf{k}_{\perp}^2$ describes the transverse spatial frequency components in order to provide variable weighting to the low-pass Lorentzian filter. While Eq. \ref{equation: paganin} applies to the two-dimensional projection images, phase retrieval can similarly be performed after CT reconstruction following 
\begin{equation}
\mu_{PR}(x, y, z) =   \mathcal{F}^{-1}\left\{\frac{\mathcal{F}[\mu_{PC}(x, y, z)]}{1 + \frac{\delta \Delta}{\mu}\textbf{k}_{3D}^2}\right\} \label{equation: 3DPR}
\end{equation}
\cite{thompson_fast_2019} where $\textbf{k}_{3D}^2 = k_z^2 + k_y^2 + k_x^2 $ , $\mu_{PC}$(x, y, z) is the CT volume of linear attenuation coefficients reconstructed from flat-fielded phase contrast projections, and $\mu_{PR}$ is the same CT volume after application of  phase retrieval. Although applying phase retrieval in 3D allows slightly higher SNR gains compared to Eq. \ref{equation: paganin}, each method is a fast, stable and effective means of achieving low-dose discrimination between low-Z monomorphous materials and their boundaries with vacuum or air \cite{kitchen_ct_2017, beltran_interface-specific_2011}. Any additional material interfaces present in the object will either be under-blurred, resulting in remnant phase contrast fringes or, in the case of higher-Z materials, over-blurred, distorting the boundary and leaving affected regions non-quantitative or completely obscured \cite{beltran_interface-specific_2011, beltran_2d_2010, croton_situ_2018}. To illustrate this, we used the projection approximation to simulate the exit surface wavefield of a low-Z object, embedded with spherical cavities and a high-Z cylinder, shown in Figure \ref{figure:projectionsims}(a). This wavefield was then propagated using the transport of intensity equation (TIE) to produce phase contrast fringes on all material boundaries (Figure \ref{figure:projectionsims}(b)). Gaussian noise was added to demonstrate the benefit of phase retrieval, which still accurately restores the low-Z cavities (Figure \ref{figure:projectionsims}(c)), although at the expense of over-blurring the high-Z interface. 

A later generalisation of Eq. \ref{equation: paganin} allows the retrieval to be specifically tuned to a specific interface within the object between two different materials \cite{beltran_2d_2010, croton_situ_2018}, introducing a second material component in $\delta_2$ and $\mu_2$ as
\begin{equation}
I(x, y) = \mathcal{F}^{-1}\left[\frac{\mathcal{F}[I_d(x, y)/I_0]}{1 + \frac{(\delta_2 - \delta_1) \Delta}{\mu_2 - \mu_1}\textbf{k}_{\perp}^2}\right] \label{equation: beltran}.
\end{equation}

\noindent This allows phase retrieval to be applied to any interface between two adjacent, non-air materials, reconstructing that edge correctly. However this approach leaves any other interfaces either over-blurred or with remnant phase contrast fringes, as shown in Figure \ref{figure:projectionsims}(d), around the air cavities. If the material properties are similar (i.e. $(\delta_2 - \delta_1) / (\mu_2 - \mu_1)$ is small), the phase contrast is weak, so the phase retrieval filter is equivalently weak and only lightly suppresses image noise.

Beltran \textit{et. al} \cite{beltran_interface-specific_2011} showed that if specific pairs of material interfaces exist in the object, a complete CT reconstruction can be comprised without over- or under-smoothed boundaries by performing separate phase-retrieved CT reconstructions for each material pair, then interleaving those reconstructions together. That technique required large spatial tolerances around high-Z components to avoid including over-blurred regions in the final stitched image. Hence, it required features of interest to be well-spatially-separated from high-Z materials, so that strong phase retrieval filtering did not spread intensity into the neighbouring feature of interest; referred to as over-blurring artefacts. Alternatively, iterative approaches aim to avoid this problem by masking out the high-Z material from projection images and using feedback cycles from the CT reconstructions to correct interface over-blurring caused by phase retrieval\cite{hehn_nonlinear_2018, hehn_propagation-based_2018}. The present paper presents an alternative method that combines 3D phase retrieval with morphological operations to prevent over-blurring artefacts from being introduced and avoids the potentially long convergence times required of iterative approaches. Hereafter, this technique will be referred to as 3D Masked Phase Retrieval (3DMPR). 

Typically, phase retrieval applications in pre-clinical studies focus on low-Z tissues where the SNR gains are highest, as seen in synchrotron studies of the breast \cite{gureyev_investigation_2014, nesterets_feasibility_2015} and lungs \cite{pollock_low-dose_2024, albers_high_2023, kitchen_ct_2017, mohammadi_quantitative_2014, dullin_functionalized_2015}, or recent small animal imaging applications on freeze-dried hearts using micro-focus sources \cite{lioliou_framework_2024}. Their mutual aim is to improve resolution and SNR while reducing radiation dose imparted to potential patients. Similar benefits have been demonstrated in brain imaging using synchrotron radiation \cite{croton_imaging_2018, croton_situ_2018}, although phase retrieval has been restricted through Eq. \ref{equation: beltran} to the bone-tissue interface as the surrounding strongly-attenuating skull can easily be over-blurred by strong phase retrieval filtering, obscuring the periphery of the brain. X-ray CT reconstructions are pertinent to medical applications for their advantages over MRI scanners such as suitability for evaluating penetrating brain injuries and acute neurological emergencies in time-critical situations \cite{temple_neuroimaging_2015, brody_chapter_2015}. Successful implementation of 3DMPR to brain imaging would open additional avenues for pursuing low-dose preclinical trials for in situ brain imaging and further secure PBI phase contrast imaging as a safe and reliable means of high speed, and high-resolution brain imaging. Additionally, phase retrieval could be made compatible with standard contrast-agent data \cite{lang_ventilationperfusion_2014, lang_increase_2016, reichmann_neodymium_2023}, expanding the flexibility of PBI imaging. 3DMPR could also be applied to the Human Organ Atlas project for regions where soft tissue meets highly attenuating bone. 

Section \ref{sec:ComputationalMethod} describes the new 3D computational process for applying strong phase retrieval to objects that contain both weakly- and highly-absorbing materials. Section \ref{sec:featurevisibilitynearinterface} shows examples of multi-material CT reconstructions performed with 3DMPR, including for polychromatic cone-beam data using a microfocus X-ray source, while providing resolution and SNR comparisons to where conventional phase retrieval is applied in 2D prior to reconstruction.

\section{Methods}
\label{sec:ComputationalMethod}
This section provides a complete description of the computational process, demonstrated using the propagation-based phase-contrast CT dataset of the head of a rabbit kitten recorded at beamline 20B2 of the SPring-8 synchrotron, Japan. Henceforth, to keep the method description general, the low-effective-Z brain-tissue, composed of atomic elements H:C:N:O:Na:P:S:Cl:K in the approximate ratio 8510:968:126:3567:7:10:5:7:6 with a density of \unit{0.986}{g.cm^{-3}} \cite{chantler_x-ray_2005, berger_xcom_2010}, will be referred to as material A. Similarly, the high-effective-Z bone surrounding the brain, H:C:N:O:Na:Mg:P:S:Ca in the ratio 3878:1483:345:3125:5:9:11:11:645 with density \unit{1.45}{g.cm^{-3}}, will be referred to as material B. When applying this method to other samples, the more attenuating material within the sample will be designated as material B.
\begin{figure}
	\centering
	\includegraphics[width = \textwidth]{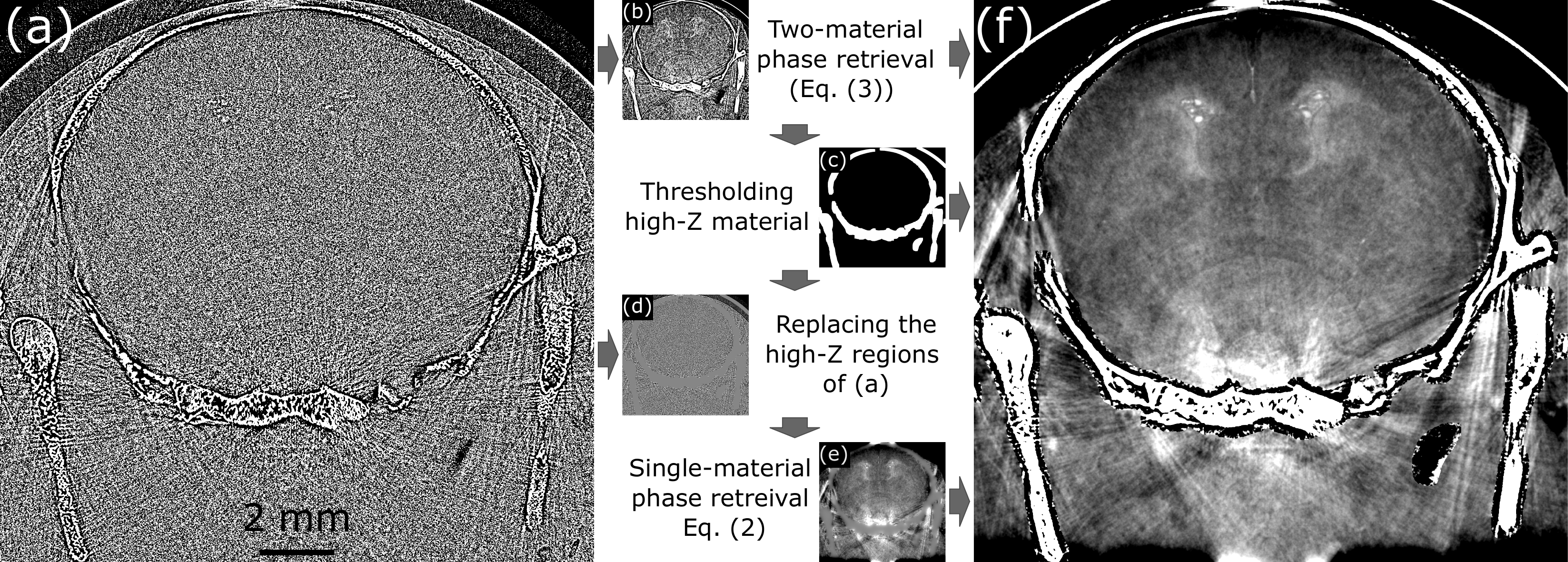}
	\caption{The masking and phase retrieval process, using a rabbit brain cross-section for demonstration. Images are labelled alphabetically according to the order in which they are computed while arrows indicate precursor images required to calculate the subsequent image.  (a) A phase-contrast CT slice without any reprocessing applied, beyond flat and dark correction. (b) The same slice phase retrieved for the bone/brain tissue interface (Eq. \ref{equation: beltran}), either achieved from 3D phase retrieval of (a) or CT reconstruction of phase retrieved projections. Upon thresholding (b), we create a pixel-wise binary mask of the high-Z material, applying a dilation filter to ensure full enclosure of the high contrast boundaries. The binary mask, (c), is then used to mask all high-Z features out of the raw phase-contrast volume by replacing them with the theoretical low-Z $\mu$ value. This creates the image in (d), which is then filtered using 3D phase retrieval of the low-Z material (Eq. \ref{equation: 3DPR}) to make (e).  Finally, the high-Z components from the two-material phase retrieved slice (b) are spliced into (e) using the binary mask, resulting in the final phase retrieved image (f), free of over-blurring artefacts from the high-Z boundaries. All non-binary images are displayed on the same colour palette which is optimised for viewing soft tissue features, leading to the bone in (f) appearing over-saturated and the subarachnoid space between the skull and brain tissue seeming indistinguishable from the surrounding agarose the object was set in.   }
	\label{figure:ComputationalMethod}
\end{figure}

Imaging was performed using synchrotron radiation at \unit{24}{\kilo e \volt}, with a \unit{5}{\metre}  propagation distance, and recorded using a 2048 × 2048 pixel Hamamatsu sCMOS camera (C11440-22C) with \unit{6.5}{\micro\metre} pixel size, fibre-optically coupled to a \unit{15}{\micro\metre} thick Gadox (Gd$_2$O$_2$S:Tb+; P43) phosphor. Due to the small width of this detector, the object was positioned with one half filling the detector and rotated through 360 degrees so that all object features are moved through the field of view. Antipodal images of the CT dataset are then stitched together, resulting in a new dataset of the whole object spinning through 180 degrees and separated by 0.1 degree increments, now with double the width. These CT datasets were reconstructed with a parallel beam geometry using filtered back projection through the TomoPy and Astra packages \cite{aarle_fast_2016, van_aarle_astra_2015} in Python, which was the language used for all analysis steps. Material parameters of the complex refractive index were taken as $\mu=$ \unit{55.1}{\meter^{-1}}, $\delta=3.93\times10^{-7}$ for material A and $\mu=$ \unit{336.83}{\meter^{-1}}, $\delta=5.43\times10^{-7}$ for material B \cite{schoonjans_xraylib_2011}. 

Figure \ref{figure:ComputationalMethod} provides a flow diagram of all the steps required to produce the phase-retrieved output, beginning with the phase-contrast volume represented by the example CT slice in Figure \ref{figure:ComputationalMethod}(a). Although all stages of the method are represented via the same CT slice, we note that 3D phase retrieval is a volume operation requiring sample volumes of dimensions at least as large as the blurring kernel. In addition to the raw phase-contrast CT volume (ie. reconstructed without phase retrieval), we require a CT volume that has been phase retrieved for the A/B material interface, using Eq. \ref{equation: beltran}. This volume, represented by Figure \ref{figure:ComputationalMethod}(b), can be calculated by phase-retrieving the phase-contrast volume in 3D \cite{thompson_fast_2019} using Eq. \ref{equation: 3DPR}, or by creating a new CT reconstruction from phase retrieved projections \cite{beltran_2d_2010}. 

The phase-retrieved volume is used to produce a binary array of the material B locations using a simple threshold that is expanded with a 3D dilation filter to ensure all voxels containing material B are included. Figure \ref{figure:ComputationalMethod}(c) shows an example binary array, calculated using a threshold of $\mu=$\unit{77.5}{\meter^{-1}} and requiring 16 iterations of a 3x3x3 kernel dilation filter due to the presence of the low absorption region (dark region) directly inside the skull. This kernel was chosen since it symmetrically expands the mask with a one pixel border; however, this could be varied as required to ensure complete coverage of all material B pixels. Using the binary array, Figure \ref{figure:ComputationalMethod}(c), allows material B to be masked out of the phase contrast volume, Figure \ref{figure:ComputationalMethod}(a), replacing the grey values of those voxels with the theoretical attenuation coefficient of material A. This results in Figure \ref{figure:ComputationalMethod}(d). Previously, we attempted to fill the masked pixels using interpolation algorithms; however, even the levels of noise in the images retrieved using Eq. \ref{equation: beltran} created too much instability, causing high contrast boundaries to appear in homogeneous sample regions. Substitution of a constant value instead minimises local contrast in and near the masked region of the image and only requires prior knowledge of the low-Z material, an assumption already required by the phase retrieval algorithms. Alternatively, the replacement attenuation coefficient could be manually measured from the phase-retrieved reconstruction (Fig. \ref{figure:ComputationalMethod}b). Next, we apply 3D phase retrieval to the masked phase contrast volume, following Eq. \ref{equation: 3DPR}, to produce the low-noise phase-retrieved CT volume in Figure \ref{figure:ComputationalMethod}(e).  Having removed any high-Z material from our sample, this retrieval cannot create any over-blurring artefacts, and performing the phase retrieval in 3D additionally provides a slight SNR boost over projection-based phase retrieval \cite{thompson_fast_2019}. 

Although 3D phase retrieval of large volume arrays can be RAM intensive, we were able to implement an out-of-core approach on a Dell Precision 5560 laptop with \unit{32}{\giga B} RAM, taking advantage of high SSD read and write speeds. Phase retrieval of a 1030$\times$1030$\times$1030 voxel data volume was achieved in just 4.6 minutes. Advanced implementation on high RAM devices/computer clusters will reduce calculation times but is generally not required. After the mask threshold and dilation settings have been selected for a particular material combination, reconstructions of similar composition could be calculated consecutively without further manual intervention, providing the possibility of rapid feedback during an experiment. 

Finally, we use the material B binary array (Figure \ref{figure:ComputationalMethod}(c)) to interleave the altered regions of the phase-retrieved volume, Figure \ref{figure:ComputationalMethod}(e), with the A/B phase-retrieved regions of Figure \ref{figure:ComputationalMethod}(b). This completes our method of 3D Masked Phase Retrieval (3DMPR), producing Figure \ref{figure:ComputationalMethod}(f), a low-noise CT volume that benefits from the strong phase retrieval designed for a low contrast material to within a few pixels of the high-Z material interface, without introducing over-blurring artefacts. This provides noise reduction within material A without eliminating features close to the A/B interface (see further examples in Section \ref{sec:featurevisibilitynearinterface}) and potentially allows new features of the low-Z material to be resolved that would have previously been obscured by noise.

\section{Results}
\label{sec:featurevisibilitynearinterface}
\begin{figure}
	\centering
	\begin{overpic}[width=0.329\columnwidth,,]{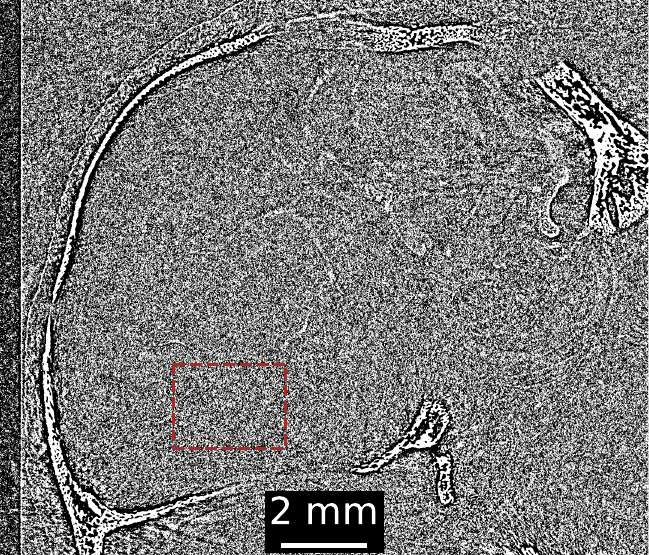}
    \put(1.0,74){\shortstack[l]{\fontsize{16}{18}\selectfont \textcolor{white}{(a)}}}
    \end{overpic}\hfill%
    \begin{overpic}[width=0.329\columnwidth,,]{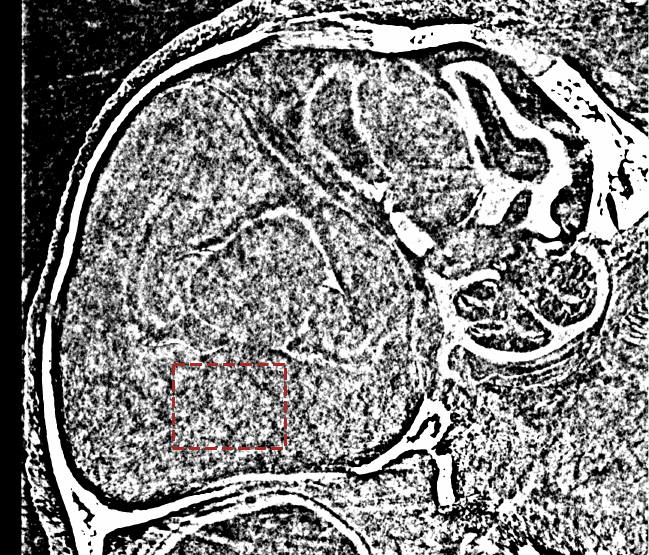}
    \put(1.0,74){\shortstack[l]{\fontsize{16}{18}\selectfont \textcolor{white}{(b)}}}
    \end{overpic}\hfill 
    \begin{overpic}[width=0.329\columnwidth,,]{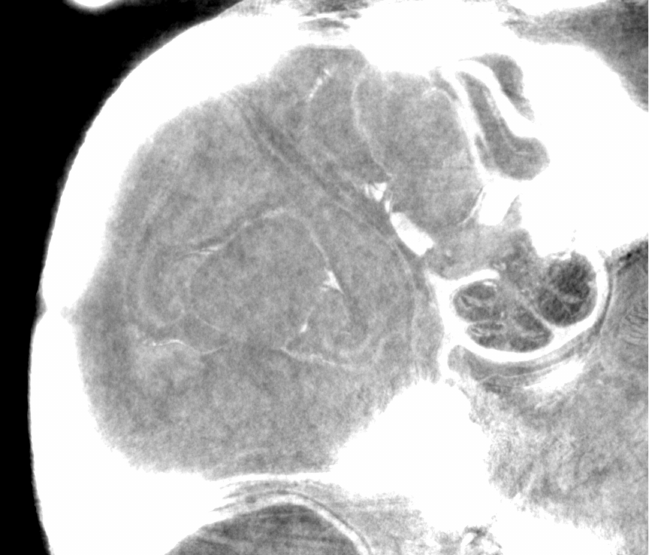}
    \put(1.0,74){\shortstack[l]{\fontsize{16}{18}\selectfont \textcolor{white}{(c)}}}
    \end{overpic}\hfill %
    \begin{overpic}[width=0.329\columnwidth,,]{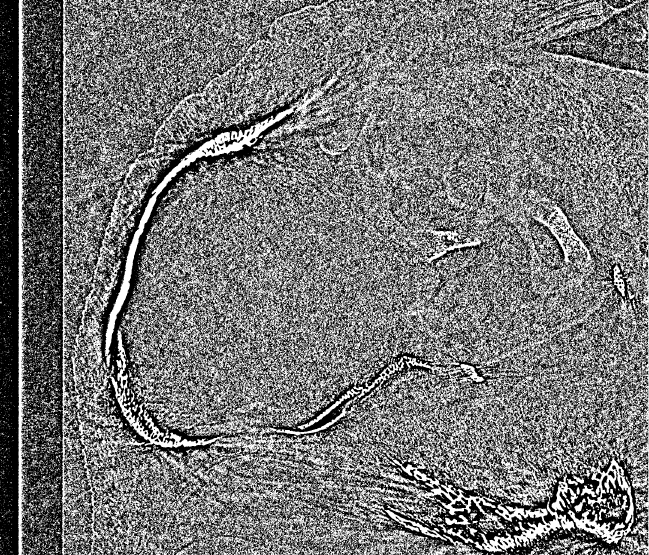}
    \put(1.0,74){\shortstack[l]{\fontsize{16}{18}\selectfont \textcolor{white}{(d)}}}
    \end{overpic}\hfill%
    \begin{overpic}[width=0.329\columnwidth,,]{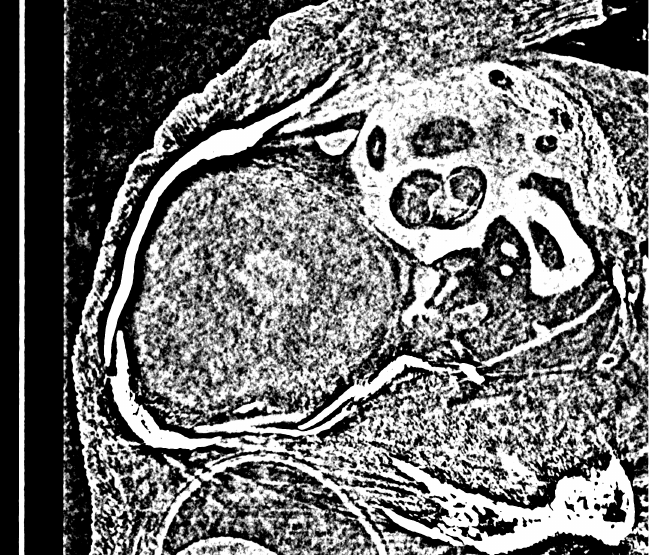}
    \put(1.0,74){\shortstack[l]{\fontsize{16}{18}\selectfont \textcolor{white}{(e)}}}
    \end{overpic}\hfill%
    \begin{overpic}[width=0.329\columnwidth,,]{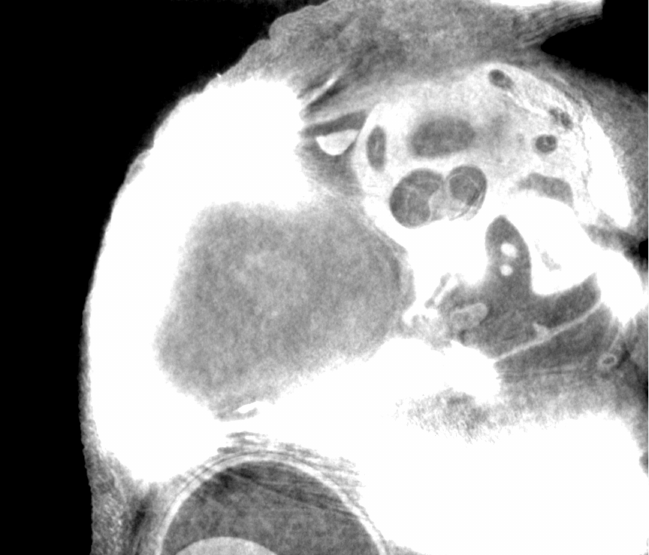}
    \put(1.0,74){\shortstack[l]{\fontsize{16}{18}\selectfont \textcolor{white}{(f)}}}
    \end{overpic}
    \begin{overpic}[width=0.499\columnwidth,,]{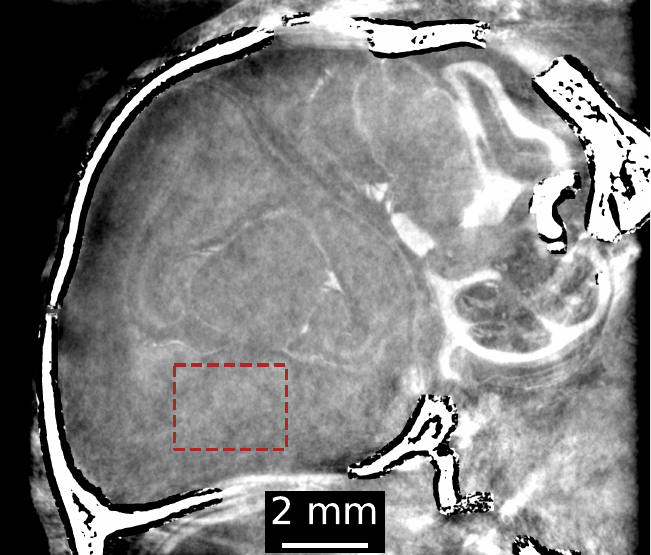}
    \put(0.5,77.5){\colorbox{black}{\shortstack[l]{\fontsize{16}{18}\selectfont \textcolor{white}{(h)}}}}
    \put(-0.5, 38){
    \begin{tikzpicture}
      \draw[white, line width = 2.0pt, ->, >=to] (-0.3, 0.45) -- (0,0);
    \end{tikzpicture}}
    \end{overpic}\hfill
    \begin{overpic}[width=0.499\columnwidth,,]{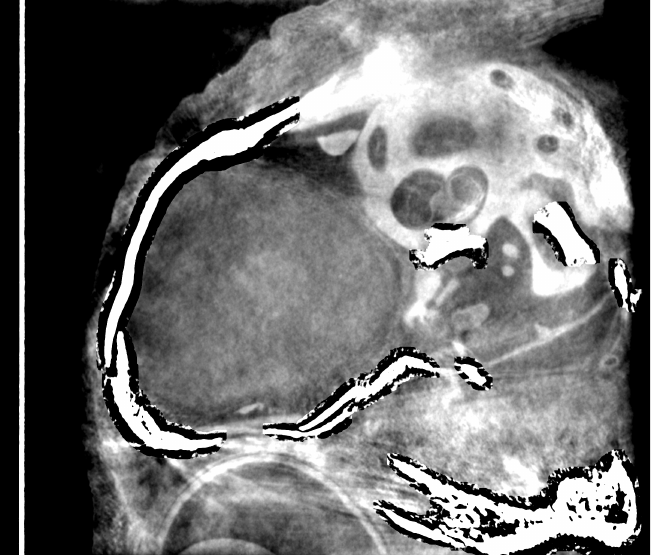}
    \put(0.5,77.5){\colorbox{black}{\shortstack[l]{\fontsize{16}{18}\selectfont \textcolor{white}{(i)}}}}
    \end{overpic} \\%
	\caption{Comparison of our masked phase retrieval method to alternative approaches, shown using sagittal cross-sections of the rabbit head CT reconstruction volumes from Section \ref{sec:featurevisibilitynearinterface}. (a) and (d) are two cross-sections of the phase contrast CT as a baseline for the other algorithms. (b) and (e) are the bone-tissue interface phase-retrieved image slices, Eq. \ref{equation: beltran}, showing considerable remnant noise. (c) and (f) shows the result of applying Eq. \ref{equation: paganin} phase retrieval to each projection under the assumption only brain tissue is present, leading to over-blurring completely obscuring brain tissue features in proximity to the bone. (h) and (i) are retrieved using 3DMPR to achieve an equally low SNR image whilst correctly reconstructing the A/B material interface. Brown boxes denote regions used for SNR analysis while the white arrow in (h) highlights a gap in the animal skull that is not visible in the Eq. \ref{equation: paganin} phase retrieval. All images use the same colour palette with the region surrounding the animal filled with agarose.}
	\label{figure:kittenbrain}
\end{figure}

To demonstrate the effectiveness of 3DMPR, we refer back to the data volume described in Section \ref{sec:ComputationalMethod}. By taking two cross-sections of the head CT volume in the sagittal plane, we compare the results of two phase retrieval approaches for the data. Figures \ref{figure:kittenbrain}(a) and \ref{figure:kittenbrain}(d) show cross sections of a standard CT reconstruction from unfiltered propagation-based phase-contrast projections, where both slices display few-to-no discernible brain features. Following the standard practice of Croton \textit{et. al} \cite{croton_situ_2018}, we can construct a new CT volume from projections phase retrieved to the A/B material interface, namely brain tissue-bone, using Eq. \ref{equation: beltran} (see Figures \ref{figure:kittenbrain}(b) and (e)). This allows resolution of some soft-tissue structures in the brain while others remain obscured by significant levels of noise still left in the reconstructions. Reducing the noise floor further requires stronger filtering, but merely applying phase retrieval for material A (Eq. \ref{equation: paganin}) in projection results in over-blurring of the high contrast interface, as seen in Figures \ref{figure:kittenbrain}(c) and (f). This overblurring leads to some features being obscured, for example the bone gap highlighted by the white arrow in Figure \ref{figure:kittenbrain}(h) which is not visible using single material phase retrieval. These overblurring artefacts may disrupt the linear attenuation coefficients far from the material boundaries, making large portions of the image no longer quantitative. The method presented in this paper, producing Figures \ref{figure:kittenbrain}(h) and (i), achieves the noise reduction of stronger filtering while also retaining edge definition in the bone-brain boundary, avoiding contrast from the bone blurring into the adjacent brain tissue. Overall, we see features of the brain anatomy with greater clarity across the entire head than with the alternative reconstruction methods.

\begin{figure}
	\centering
    \hfill
    \begin{overpic}[width=0.32\columnwidth,,]{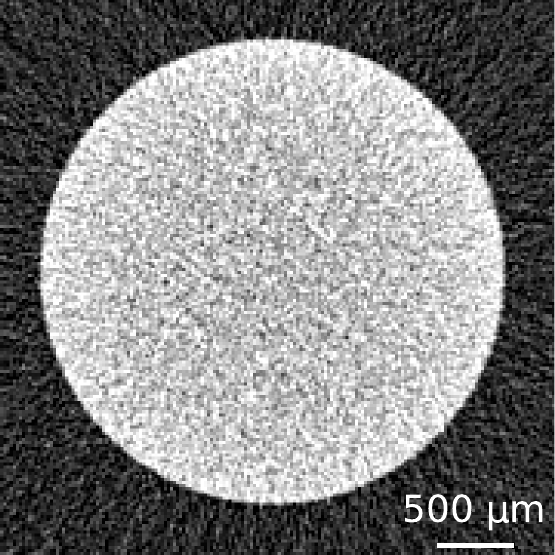}
    \put(1.0,87.5){\shortstack[l]{\fontsize{16}{18}\selectfont \textcolor{white}{(a)}}}
    \end{overpic}
	\begin{overpic}[width=0.32\columnwidth,,]{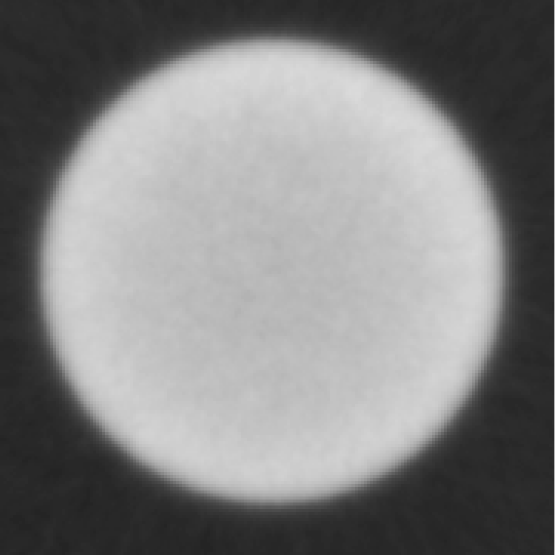}
    \put(1.0,87.5){\shortstack[l]{\fontsize{16}{18}\selectfont \textcolor{white}{(b)}}}
    \end{overpic}
	\begin{overpic}[width=0.32\columnwidth,,]{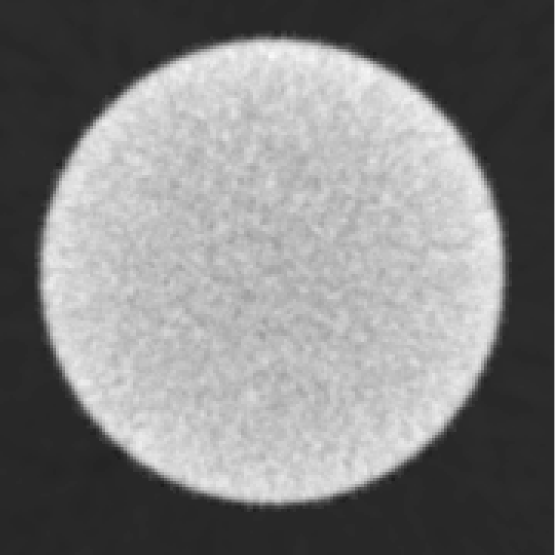}
    \put(1.0,87.5){\shortstack[l]{\fontsize{16}{18}\selectfont \textcolor{white}{(c)}}}
    \end{overpic}\hfill\vspace{0.5em}\newline%
    \begin{overpic}[width=0.72\columnwidth,,]{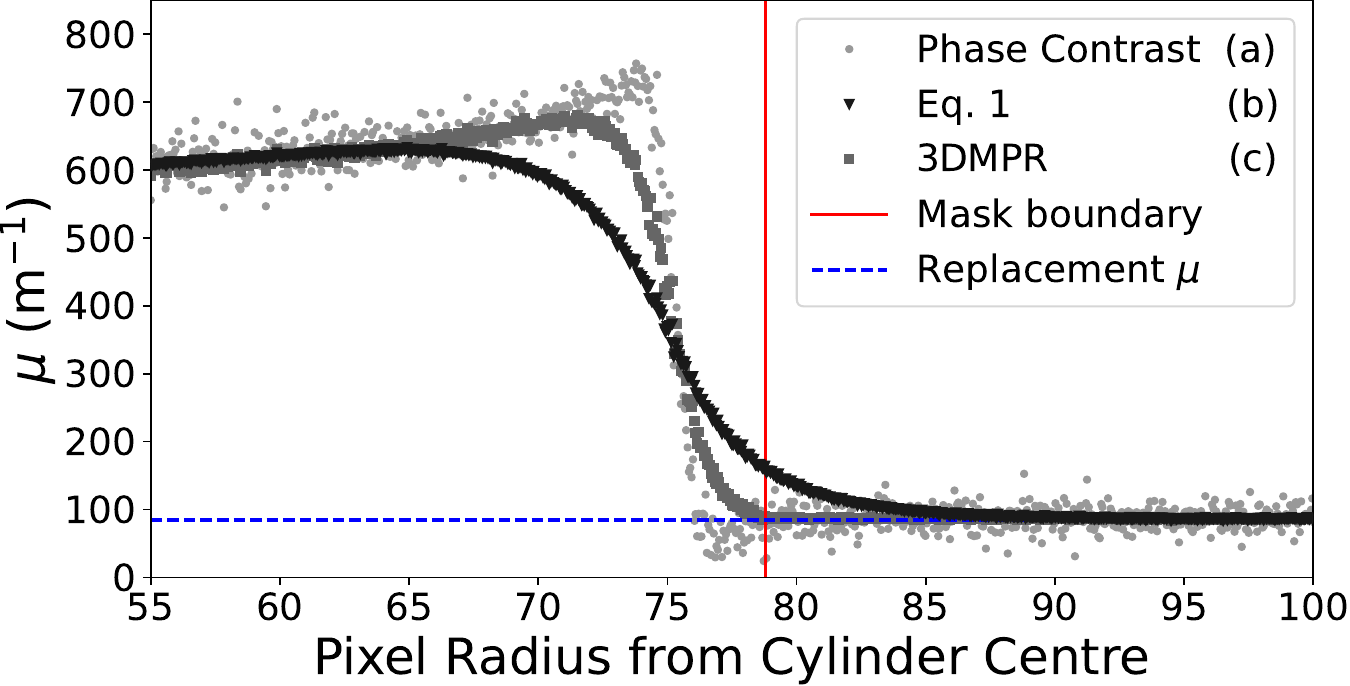}
    \put(-3,45.0){\shortstack[l]{\fontsize{16}{18}\selectfont \textcolor{black}{(d)}}}
    \end{overpic}\vspace{0.5em}
	\caption{Quantification of edge resolution in a \unit{3}{mm} thick aluminium rod submerged in water after phase retrieval was applied to material A, water. (a) shows a phase contrast slice of the aluminium cylinder while (b) and (c) show the same slice after from phase retrievals performed through the single-material algorithm (Eq. \ref{equation: paganin}) and 3DMPR, respectively. (d) plots line profiles produced by azimuthally averaging around the entire aluminium-water interface to compare the raw phase contrast resolution to that of the conventional 2D phase retrieval and our 3D masking method. A solid vertical line is used to show the mask radius, notably not visible in (c), while a horizontal line shows the theoretical $\mu$ value of the low-Z material (water) used for replacing the masked Aluminium in the 3D reconstruction. Images (b) and (c) share the same colour palette. }
	\label{figure:PMMA-AlPhantom}
\end{figure}

\subsection{SNR Characterisation}
SNR characterisations were performed using the regions indicated in Figures \ref{figure:kittenbrain}(a), (b) and (h). Values were calculated in a 5-slice radius, using the standard deviation as the error. Raw phase contrast reconstructions (Figure \ref{figure:kittenbrain}(a)) produced a SNR of $1.09\pm0.02$, which was improved under two-material phase retrieval (Equation \ref{equation: beltran}) interface to $37.0\pm0.3$. 3DMPR further increases SNR by a factor of 6.8 to $252\pm6$, representing a total SNR gain of 231, without compromising resolution. However, the region being analysed includes variations due to features of the brain tissue as shown in Figure \ref{figure:kittenbrain}(h) which are recorded as noise measured through the standard deviation. The SNR gain of 6.8 is then likely an underestimation, but still represents a further potential dose reduction of $6.8^2 \approx 46$-fold \cite{kitchen_ct_2017} that may become part of achieving lower-dose brain imaging. 

    \subsection{Boundary Preservation}
To quantify the method's ability to preserve high contrast material boundaries and avoid over-blurring, we explore the simple and well-characterised dataset of an aluminium pin submerged in water. This dataset was recorded on a \unit{50}{\micro\meter} pixel Hamamatsu CMOS flat panel detector (C9728DK-10) using a microfocus X-ray source (THE-Plus from X-RAY WorX Gmbh) with a silver transmission target at a tube voltage of \unit{35}{\kilo V_p}. The mean X-ray energy was determined by comparing $\mu$ values of water measured in CT at the depth of the aluminium pin, resulting in a mean energy of \unit{19.58}{\kilo e \volt}. The source-to-object distance was \unit{0.96}{\meter} with the source-to-detector distance set to \unit{2.40}{\meter}, resulting in a 2.50 times magnification and effective propagation distance of \unit{0.576}{\meter}. CTs were reconstructed with 3,271 projections at 0.11 degree increments using filtered back projection in a fan beam geometry\cite{aarle_fast_2016, van_aarle_astra_2015}. For this sample, material A is taken to be water, $\mu=$ \unit{84.72}{\meter^{-1}} and $\delta=6.00\times10^{-7}$, while material B is the comparatively high-Z aluminium with parameters $\mu=$ \unit{985.86}{\meter^{-1}} and $\delta=1.38\times10^{-6}$ \cite{schoonjans_xraylib_2011}. 

Figure \ref{figure:PMMA-AlPhantom} shows example portions of a CT slice through the aluminium pin for phase retrieval (b) with Eq. \ref{equation: paganin} and (c) with our 3DMPR method. Each figure applies the same filtering to regions surrounding the aluminium cylinder, but (c) uses two-material phase retrieval to more accurately preserve the A/B material boundary. For 3DMPR, the aluminium cylinder was masked using a $\mu$ threshold of \unit{300}{\metre^{-1}} with two iterations of a 3$\times$3$\times$3 dilation filter. Despite the aluminium insert having uniform density, both figures \ref{figure:PMMA-AlPhantom}(b) and (c) show a decrease in attenuation coefficient toward the centre, which is a cupping artefact caused by beam hardening \cite{barrett_artifacts_2004}. This results from variable penetration depths of the polychromatic X-ray spectrum \cite{croton_situ_2018}. Figure \ref{figure:PMMA-AlPhantom}(d) plots azimuthally-averaged line profiles across the A/B boundary  in (a), (b) and (c), incorporating the phase contrast profile from the non-phase-retrieved CT slice for comparison. A solid vertical line represents the edge of the mask used for replacing material B. The horizontal dashed line denotes the theoretical linear attenuation coefficient value used for the temporary high-Z replacement, calculated using the mean energy of the polychromatic source, which closely resembles the measured linear attenuation coefficient outside the region being replaced. Figure \ref{figure:PMMA-AlPhantom}(d) shows the edge profile of our 3DMPR approach with dark grey datapoints. This more closely resembles the true boundary interface of the aluminium pin, whereas the Eq. \ref{equation: paganin} phase retrieval approach, shown in Fig. \ref{figure:PMMA-AlPhantom}(d) with the black datapoints, produces a low-contrast, blurred edge due to over-blurring, even when retrieving at a relatively small effective propagation distance of \unit{0.576}{\meter}. This over-blurring leaves the area around the boundary unsuitable for quantitative analysis, even for any features of interest not obscured by the edge blurring. To measure the blurring extent of each approach, we numerically differentiated each edge profile and applied a Pearson VII fit to determine the full width at half maxima of the effective point spread function (PSF) \cite{croton_situ_2018}. Pearson VII functions are appropriate fits for PSFs \cite{croton_situ_2018} since they have freedom that enables their shape to lie on a spectrum between Lorentzian and Gaussian. Through this measure, the spatial resolution of the conventional phase-retrieval approach was found to be \unit{108\pm2}{\micro\meter}, while our approach was measured at \unit{54\pm1}{\micro\meter}. This demonstrates a clear improvement in material boundary definition while retaining the same SNR boost in the low-Z material of $4.2\times$, measured using the same method described in Section 3A. This SNR boost will be particularly valuable in improving weakly-attenuating feature resolution in low flux imaging scenarios. 
\begin{figure}
	\centering
	\hspace{-0.6mm}\begin{overpic}[width=0.16\columnwidth,,]{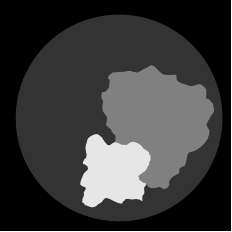}
    \put(1,85){\shortstack[l]{\fontsize{11}{16}\selectfont \textcolor{white}{(a)}}}%
    \end{overpic}\hspace{0.5em}%
	\begin{overpic}[width=0.33\columnwidth, height = 0.16\columnwidth,,]{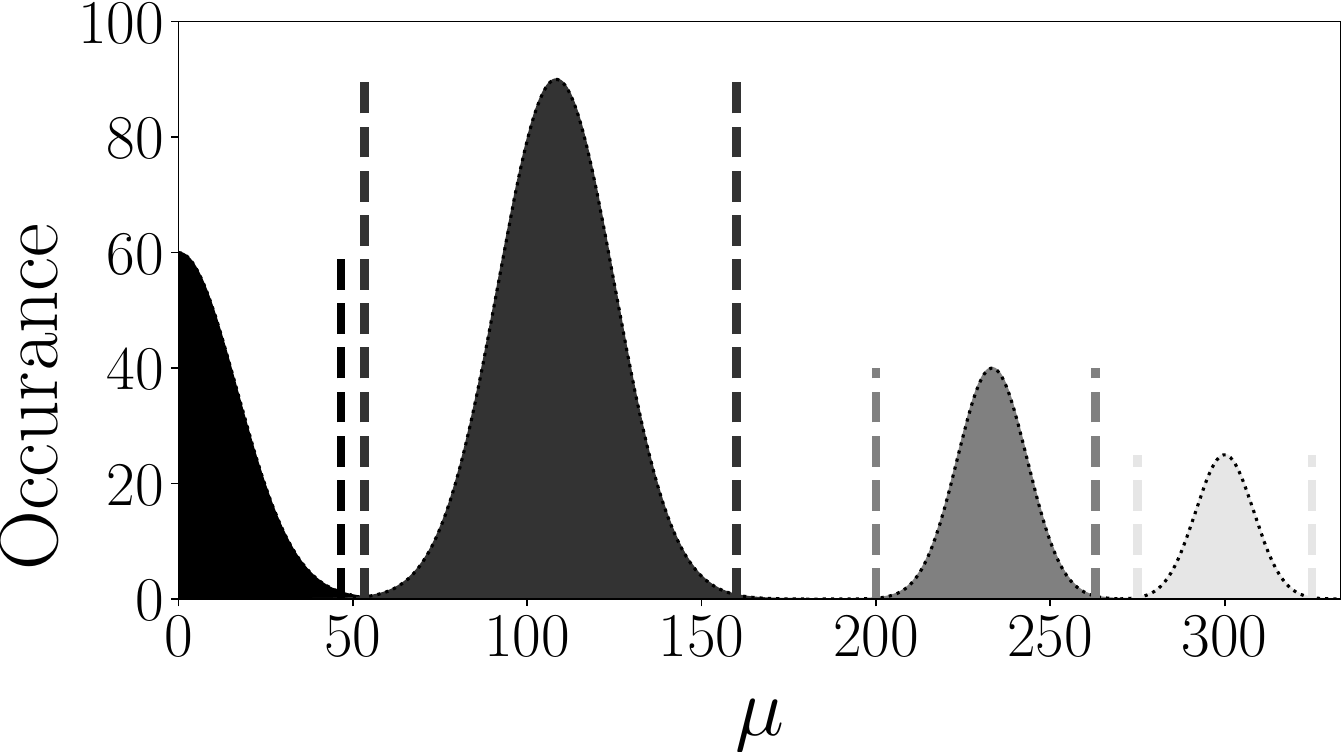}
    \put(-4.5,41.5){\shortstack[l]{\fontsize{11}{16}\selectfont \textcolor{black}{(b)}}}
    \put(31,35){\shortstack[l]{\fontsize{13}{15}\selectfont \textcolor{black}{I}}}
    \put(68,27){\shortstack[l]{\fontsize{13}{15}\selectfont \textcolor{black}{II}}}
    \put(86,21){\shortstack[l]{\fontsize{13}{15}\selectfont \textcolor{black}{III}}}
    \end{overpic} \\%

    \includegraphics[width=0.45\columnwidth]{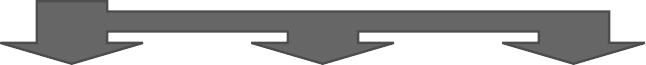}\hspace{0.9cm}\\
	
	\begin{overpic}[width=0.16\columnwidth,,]{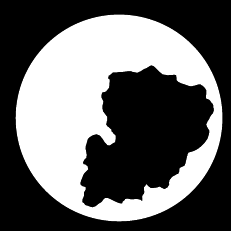}
    \put(1,85){\shortstack[l]{\fontsize{11}{16}\selectfont \textcolor{white}{(c)}}}
    \end{overpic}\hspace{0.5em}%
	\begin{overpic}[width=0.16\columnwidth,,]{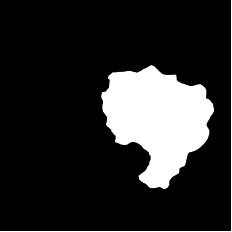}
    \put(1,85){\shortstack[l]{\fontsize{11}{16}\selectfont \textcolor{white}{(d)}}}
    \end{overpic}\hspace{0.5em}%
    \begin{overpic}[width=0.16\columnwidth,,]{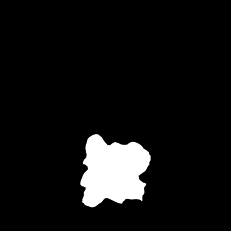}
    \put(1,85){\shortstack[l]{\fontsize{11}{16}\selectfont \textcolor{white}{(e)}}}
    \end{overpic}\\%

    \vspace{-0.075cm}
    \includegraphics[width=0.10\columnwidth, height=0.025\columnwidth]{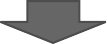}\hspace{0.9cm}
    \includegraphics[width=0.10\columnwidth, height=0.025\columnwidth]{gui5004fig5l.pdf}\hspace{0.9cm}
    \includegraphics[width=0.10\columnwidth, height=0.025\columnwidth]{gui5004fig5l.pdf}\\
    
    \begin{overpic}[width=0.16\columnwidth,,]{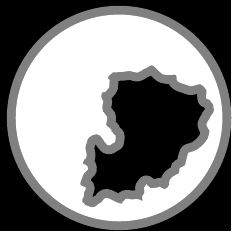}
    \put(1,85){\shortstack[l]{\fontsize{11}{16}\selectfont \textcolor{white}{(f)}}}
    \end{overpic}\hspace{0.5em}%
    \begin{overpic}[width=0.16\columnwidth,,]{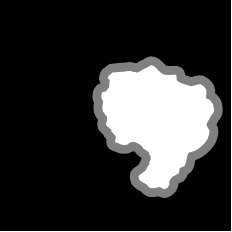}
    \put(1,85){\shortstack[l]{\fontsize{11}{16}\selectfont \textcolor{white}{(g)}}}
    \end{overpic}\hspace{0.5em}%
    \begin{overpic}[width=0.16\columnwidth,,]{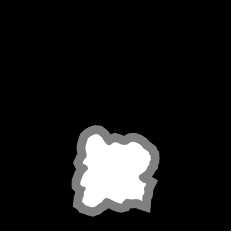}
    \put(1,85){\shortstack[l]{\fontsize{11}{16}\selectfont \textcolor{white}{(h)}}}
    \end{overpic}\\%
    
    \vspace{-0.075cm}
    \includegraphics[width=0.10\columnwidth, height=0.025\columnwidth]{gui5004fig5l.pdf}\hspace{0.9cm}
    \includegraphics[width=0.10\columnwidth, height=0.025\columnwidth]{gui5004fig5l.pdf}\\%

    \hspace{0.18\columnwidth}
    \begin{overpic}[width=0.16\columnwidth,,]{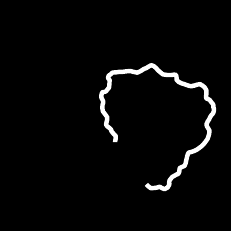}
    \put(1,85){\shortstack[l]{\fontsize{11}{16}\selectfont \textcolor{white}{(i)}}}
    \end{overpic}\hspace{0.5em}%
    \begin{overpic}[width=0.16\columnwidth,,]{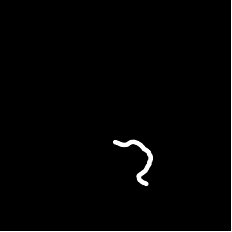}
    \put(1,85){\shortstack[l]{\fontsize{11}{16}\selectfont \textcolor{white}{(j)}}}
    \end{overpic}
    \hspace{0.16\columnwidth}
	\caption{Conceptualised application of 3DMPR to a multi-material object composed of materials I,II and III, as shown in (a). Images shown represent single slices of the CT volume with the assumption of contrast isolation between all materials as in the histogram (b). Dashed, vertical lines demonstrate example threshold bounds for creating material masks, shown in (c)(d)(e). The inverse of these masks, black regions, can be used to mask away all other material components, allowing phase retrieval to be performed to internal material regions using Eq. \ref{equation: paganin}. The boundary regions are interleaved from appropriately tuned retrievals, located using binary operations between the material masks after an additional dilation (f)(g)(h). (i) shows the result of a binary 'and' operation between the material I and II masks, creating a trace of the I/II interface for interleaving. (j) demonstrates the same for the II/III interface, overcoming the convention limit of two-way material boundaries. }
	\label{figure:generalisation}
\end{figure}

\subsection{Expansion to 3+ Material Interfaces}
While section \ref{sec:ComputationalMethod} demonstrated the 3DMPR process using a two-material sample, here we generalise to higher sample compositions and demonstrate how masked phase retrieval may be used to overcome a restriction of the two-material algorithm of Eq. \ref{equation: beltran}. The phase-retrieval algorithms of Eq. \ref{equation: paganin} and Eq. \ref{equation: beltran} require that each material is only in contact with a maximum of one other material or vacuum. Otherwise, only parts of the material's boundary will be quantitatively reconstructed.   

Phantoms containing relevant information in multiple attenuation levels adjacent to high-Z boundaries were not imaged during experimentation; therefore, to demonstrate the concept we consider a mock 3-material CT slice, shown in Figure \ref{figure:generalisation}(a), composed of materials I, II, and III and possessing 4 unique interfaces: I/air, I/II, I/III and II/III. Phase retrieval with the two-material algorithm (Eq. \ref{equation: beltran}) cannot fully resolve the boundary of any material. For example, phase retrieval focused on the II/I interface will leave the II/III interface non-quantitative. Our approach relaxes the requirement for spatially-isolated materials but does require materials to be contrast-resolved, as in the histogram of Figure \ref{figure:generalisation}(b). This allows two-sided thresholding to create binary masks for each material, Figures \ref{figure:generalisation}(c,d,e), in place of the single-sided thresholding used in Section \ref{sec:ComputationalMethod}. Although the material distributions in Figure \ref{figure:generalisation}(b) may appear idealised, 3DMPR is proposed to counter over-blurring from phase retrieval specifically where high-Z materials are present in a comparably low-Z medium. Otherwise, materials with overlapping $\mu$ distributions may share equivalent enough phase characteristics to be considered a single material, allowing an average value to be used in conventional TIE phase retrieval and negating the need for 3DMPR. Alternatively, other mask segmentation methods may be employed, such as region-growing algorithms or other advanced processes that utilise position information and that would benefit from phase-contrast fringes highlighting sample boundaries, particularly in the presence of noise.

3DMPR is then divided into two categories; the relatively constant regions within each material, termed homogeneous regions, and regions around material boundaries containing two materials, referred to as heterogeneous. Phase retrieval for homogeneous regions is performed to the CT volume, using equation Eq. \ref{equation: paganin}, after replacing all other image components with the theoretical material value using the inverse of each mask (Figure \ref{figure:generalisation}(c), (d) and (e)). A dilation filter may be required on the inverse masks to ensure heterogeneous regions are not included. When combined, these reconstructions provide strong noise suppression within each material, without over-blurring low-Z components, but do not yet include the material boundaries. These heterogeneous sample regions must be interleaved from appropriately tuned reconstructions, requiring knowledge of the boundary locations, acquired through further manipulation of the binary masks (Figures \ref{figure:generalisation}(c, d, e)). The masks are expanded by an equivalent amount via dilation filters  (Figures \ref{figure:generalisation}(f, g, h)) until masks of adjacent materials overlap. Binary `and' operations are then used to isolate the overlapping regions which will be centred about the material boundaries. Figures \ref{figure:generalisation}(i) and (j) show examples of this for the II-I and II-III material interfaces, respectively, which can then be used to select these regions from the appropriate reconstructions.

Interleaving all homogeneous and heterogeneous regions together, and iterating the method proven in Sections 2, 3a and 3b, will produce a noise suppressed CT reconstruction with quantitative resolution at all material boundaries.

\section{Discussion}
3DMPR is a method for interface-specific phase retrieval which overcomes filtering limitations at the boundaries of high-Z materials while maintaining the same assumptions as Eq. \ref{equation: paganin}, i.e. prior knowledge of the sample composition and resolution of phase contrast fringes. Note that if the materials are not known, their properties can nevertheless be determined by analysing the phase contrast fringes in the CT \cite{alloo_tomographic_2022}. Although clear material distinctions in CT are required to create the 3D mask, this is a defining aspect of the posed problem of applying phase retrieval near highly attenuating objects. If such material distinctions are not present then regular phase retrieval may be applied instead.  Applying 3DMPR provides greater high frequency noise suppression than Eq. \ref{equation: beltran} alone, increasing the clarity of all features including any artefacts already present in the CT reconstruction. This is shown in Figure \ref{figure:ComputationalMethod} where streak artefacts \cite{barrett_artifacts_2004} are visible from photon trajectories parallel to the edge of the bones. Because the streak artefacts arise during the CT reconstruction step, and not the phase retrieval step, the approach described in this paper will not address these artefacts. Similar to the Paganin and Beltran algorithms, 3DMPR will provide the greatest benefits for large propagation distances, low-z materials and low energies, subject to conditions satisfying the TIE. This includes requiring sufficiently small pixel sizes to resolve the phase contrast fringes. Specifically, if $(\delta_2 - \delta_1)/(\mu_2 - \mu_1)$ is smaller than $\delta/\mu$ used in 3DMPR then the SNR will improve.

We believe 3DMPR will increase the application scope of phase retrieval, by improving results in the presence of highly-attenuating materials, such as the medical bone-flesh examples we present in figures \ref{figure:ComputationalMethod} and \ref{figure:kittenbrain}. More complex extensions with 3-materials could add a distinction for the cartilage seen in figures \ref{figure:kittenbrain}(b) and (e). Similar medical applications could include distinguishing soft material from highly-attenuating implants/contrast agents in and around bone where three materials may meet, creating a scenario where standard phase retrieval has difficulties  and neutron imaging may otherwise be required \cite{isaksson_neutron_2017}. Other examples of complex sample applications could be analysing soil samples or manufactured objects, in particular with metals or minerals. Each phantom would require an initial optimisation of the masking process. For example, see the difference in dilation filters between the datasets of Figures \ref{figure:kittenbrain} and \ref{figure:PMMA-AlPhantom} where the former required many more dilations to cover the low attenuation region between the skull and brain tissue, known as the subarachnoid space, when starting from the bone threshold as an initial seed. However, the processing of similar samples would follow the same settings, which could be used to establish an efficient workflow with no further manual interventions required.
 
The problem of phase retrieval with two-material samples has also been tackled before using successive retrieval on the one volume \cite{ullherr_correcting_2015}, including presenting what we believe is the first three dimensional phase retrieval filter and a similar method to ours for retrieving near high-Z materials \cite{ullherr_correcting_2015}. The primary differences between each method are that Ullherr et. al \cite{ullherr_correcting_2015} apply the second phase retrieval directly to the Eq. \ref{equation: beltran} phase retrieved CT and that they do not perform material replacement in the masked regions. Applying 3DMPR to the two-material retrieved CT volume, instead of a phase contrast volume, may provide a slightly higher SNR due to phase retrieval being more noise tolerant than phase contrast CT. However, replacing the mask area creates an effectively homogeneous material, eliminating over blurring from neighbouring regions or erosion of features due to neighbouring masked regions without the need for additional post-processing. Haggmark et al. further  explore the Ullherr et. al approach \cite{ullherr_correcting_2015} alongside two-material phase-retrieval \cite{beltran_interface-specific_2011} using the labels `linear' and `parallel', respectively \cite{haggmark_comparison_2017}. 

\section{Conclusion}
\label{sec:conclusion}
We present a method for achieving strong noise reduction during phase retrieval of multi-material objects, without creating over-blurring along material boundaries with high-Z materials. By using simple thresholding techniques for material isolation in CT, the method is computationally efficient and compatible with polychromatic reconstructions. A rabbit kitten brain CT was used to demonstrate the algorithm and its ability to improve image quality in complex biological samples. SNR values increased from $1.08\pm0.02$ in the phase contrast images to $252\pm6$ using 3DMPR, an increase of 6.8 times over the two-material phase retrieval algorithm, whilst preserving the same boundary definition between the high and low-Z materials.  Conversely, when applied to a separate, well-characterised dataset, edge resolution across the high-Z boundary was markedly improved from \unit{108\pm2}{\micro\meter} using the approach in Eq. \ref{equation: paganin} to \unit{54\pm1}{\micro\meter} with the new approach. This method will allow strong SNR-boosting phase retrieval to be applied to a wider range of multi-material, including those with three or more material interfaces. 

\section{Acknowledgements}
This experiment used rabbit kittens that had been used in experiments conducted with approval from the SPring-8 Animal Care (Japan) and Monash University (Australia) Animal Ethics Committees. J.A.P is supported by a Research Training Program (RTP) Scholarship and the J. L. Williams Top-up Scholarship. K.M. acknowledges support from the Australian Research Council (FT18010037). S.B.H. is an NHMRC Principal Research Fellow. This work was funded by NHMRC 2021 Ideas Grants Application 2012257 and supported by the Victorian Government's Operational Infrastructure Support Program.

\section{Data Availability Statement}
The datasets and/or analysis pipelines used for this manuscript are available from the corresponding author on reasonable request.

\bibliographystyle{unsrt}

\end{document}